\documentclass[aps,prb,twocolumn,groupedaddress]{revtex4-1}

\usepackage{amsmath}
\usepackage{graphicx}
\usepackage{color}

\begin{document}

\title{Tightly bound excitons in two-dimensional semiconductors with a flat valence band}

\author{Maxim Trushin}
\affiliation{Centre for Advanced 2D Materials, National University of Singapore, 6 Science Drive 2, 117546, Singapore}

\date{\today}

\begin{abstract}
This theoretical paper offers an explicit expression for the binding energy
of excitons in a two-dimensional semiconductor with a flat valence band.
The formula has been derived quasiclassically assuming that the exciton is tightly bound;
i.e., its ground-state radius is determined by the intrinsic polarizability of the semiconductor
rather than by the dielectric properties of the environment.
The model is relevant to a few two-dimensional semiconductors discovered recently,
including distorted 1T-TiSe$_2$, with a supposedly unstable electronic ground state.
The valence band {\em flatness} reduces the exciton binding energy,
which also may have an effect on the phase transition to an excitonic insulator.
\end{abstract}

%\pacs{72.20.Dp,72.80.Vp}
%\keywords{}

\maketitle

\section{Introduction}

Two-dimensional (2D) semiconductors demonstrate optical and electronic properties
qualitatively different from their three-dimensional (3D) parent crystals
due to the crystallographic symmetry changes,
ultimately strong electron confinement within a 2D plane, 
and absence of internal volume resulting in a great exposure to the environment.
\cite{Science2016review,CSR2014atlas,Nanoscale2011review}
The latter offers an interesting opportunity to tune Coulomb interactions inside
2D semiconductors just by changing the surrounding media.\cite{PRB2012-conf-env,lin2014dielectric,kylanpaa2015binding,stier2016probing}
The simplest way to see this effect is to measure the optical absorption governed by excitons ---
the electron-hole (e-h) pairs bound together by Coulomb forces.\cite{wang2018colloquium}
The exciton binding energy determines
the strongest peak in the absorption spectra. 
Since the dielectric screening in 2D semiconductors can be greatly reduced,
the exciton binding energy may reach hundreds of meV.
\cite{NanoLett2015hill,PRL2014chernikov,PRL2014he,SSC2015hanbicki,SR2015zhu,NatMat2014ugeda}
Despite having such high binding energy, the corresponding quasiclassical radius of the ground-state orbital
\cite{stier2016probing,stier2016exciton}
remains about 1 nm --- still substantially larger than the lattice constant of the order of 1 $\mathring{\mathrm{A}}$.
Thus, such excitons can be regarded as Wannier-Mott excitons, sometimes referred to
as {\em tightly bound} 2D Wannier-Mott excitons\cite{PRL2014he} to emphasize the strong binding of the e-h pairs involved.

Relying on the effective-mass approximation and hydrogen-like model,  
the exciton spectrum can be written as a conventional 2D Rydberg series given by
$E_n= -E_b/(2n+1)^2$, $n=0,1,2...$, where the binding (ground state) energy $E_b=|E_{n=0}|$ reads
\begin{equation}
 \label{binding00}
 E_b=\frac{2\hbar^2}{\mu a_B^2}.
\end{equation}
Here, $a_B=\epsilon_\mathrm{env}\hbar^2/(\mu e^2)$ is the effective Bohr radius, 
$\epsilon_\mathrm{env}$ is the relative dielectric permittivity of the environment, $e$ is the elementary charge,
$\mu$ is the reduced e-h mass, and $\hbar$ is the Planck constant.
Equation (\ref{binding00}) shows that the binding energy rapidly increases when screening is reduced ($\epsilon_\mathrm{env}\to 1$),
and the excitons become tightly bound (small $a_B$).
One can imagine a set of parameters when the binding energy exceeds the fundamental band gap
that leads to the ground-state reconstruction,\cite{PR1967ground-state-EI,keldysh1968collective}
and the semiconductor becomes a so-called {\em excitonic insulator}
predicted a half a century ago.\cite{keldysh1965possible,CLOIZEAUX1965259}
Much later, the experimental evidence of this phenomenon was found
in several bulk semiconductors.\cite{PRL1991bucher,PRL2007cercellier,PRL2009wakisaka,lu2017zero,PRL2017mor}
The advent of 2D semiconductors and heterostructures reignited the interest of experimentalists\cite{du2017evidence} and 
theoreticians\cite{BRUNETTI2019,PRB2018jiang} in 2D excitonic insulators.
Inspired by the recent advances in the field we address the relevant question: What limits
the binding energy of a tightly bound 2D Wannier-Mott exciton from above?

\begin{figure}
\includegraphics[width=\columnwidth]{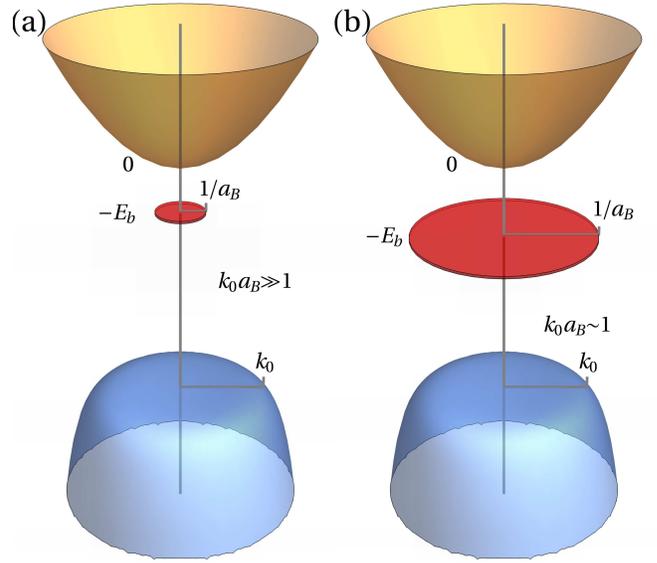}
\caption{\label{fig1} The valence band dispersion contains a term proportional to $(k/k_0)^p$, $p>>2$,
where $k_0$ is the typical radius of the flat region in the momentum space.
The conduction band dispersion remains parabolic and allows for an effective-mass description.
The exciton is characterized by the binding energy $E_b$ and the effective Bohr radius $a_B$.
(a) A weakly bound exciton corresponds to the case $a_B k_0 \gg 1$ when the effective-mass approximation applies.
(b) Once $a_B k_0 \sim 1$ the effective-mass approximation fails for the valence band
and the nonparabolic term must be taken into account.}
\end{figure}

The first limitation is obvious: Once the exciton radius $a_B$ becomes comparable
to the lattice constant the very Wannier-Mott picture becomes inapplicable,
setting the minimal $a_B$ to be about a few  $\mathring{\mathrm{A}}$. Assuming a typical $\mu$ to be of the order of
$0.1m_0$ ($m_0$ is the free electron mass) we obtain the maximum $E_b$ well above 1 eV.
Such high binding energy has never been detected in 2D semiconductors despite using different substrates
to reduce the screening.\cite{stier2016probing,NatMat2014ugeda}
The second limitation is less obvious but also well known: 
The interaction forming an e-h pair in the 2D limit is not Coulomb-like, $-1/r$,
but rather logarithmic, $\ln(r/r_s)$, where $r_s>r$ is the screening length determined by
the intrinsic polarizability of a semiconducting layer.\cite{cudazzo2011dielectric}
Hence, the exciton binding energy in a polarizable semiconductor diverges
slower than $1/a_B^2$ initially suggested by Eq. (\ref{binding00}).
The upper limit for the binding energy of a tightly bound 2D Wannier-Mott exciton is therefore 
a few hundreds of meV for typical $r_s$ value of about 5 nm.\cite{berkelbach2013theory}

There is, however, the third limitation that we are now focusing on.
The effective-mass approximation is applicable only in the very vicinity of the band extrema.
Once the exciton radius $a_B$ becomes so small, that the states with momenta 
beyond this vicinity start contributing to the exciton ground state, the effective-mass approximation obviously fails.
To be specific, we define a characteristic wave vector, $k_0$, that separates
the states with $k<k_0$, where the effective-mass approximation is still applicable,
from the states with $k>k_0$, where the band dispersion becomes nonparabolic.
Figure \ref{fig1} demonstrates a simplified version of such a band structure.
The valence band is shown to be relatively flat, so that the reduced effective mass is mostly
determined by the conduction band as long as $k<k_0$. At $k>k_0$ the valence band dispersion
drops much faster than $-k^2$, which drastically changes the exciton binding energy once
$a_B^{-1}$ becomes comparable with $k_0$.

In what follows, we derive the exciton binding energy quantizing e-h motion quasiclassically.
We explicitly include all three parameters discussed above: $a_B$, $r_s$, and $k_0$. 
The result given by Eq. (\ref{binding11}) is relevant, in particular, 
for distorted 2D 1T-TiSe$_2$, where $k_0\sqrt{a_Br_s/2}\sim 1$.

\section{Excitonic bound states}

To derive an explicit expression for the exciton spectrum we employ 
the Bohr-Sommerfeld quantization rule known to be applicable to dynamical systems in which trajectories are closed,
no matter which dispersion the involved particles obey.\cite{Granovskii,PRL2007shytov,PRA1994shvartsman,PRA2015shvartsman}
An e-h pair is assumed to be created by absorbing a single photon thanks to the direct interband optically-allowed transition. 
An electron and a hole are moving in opposite directions with the same absolute value of their momenta,
$k_e=k_h=k$, so that the exciton does not move as a whole.
The electron energy then reads $E_e=\frac{\hbar^2 k^2}{2m_e}$, where $m_e$ is the effective electron mass, and
the hole energy is given by $E_h=\frac{\hbar^2 k^2}{2m_h} + \beta \left|\frac{k}{k_0}\right|^p$, where
$m_h$ is the effective hole mass, and $\beta$, $k_0$, $p\gg 2$ are the valence band parameters
describing its {\it flatness}. The total energy of the e-h relative motion reads
\begin{equation}
 \label{total}
 E=\frac{\hbar^2 k^2}{2\mu} + \beta \left|\frac{k}{k_0}\right|^p - V(r),
\end{equation}
where $\mu=m_e m_h/(m_e+m_h)$ is the reduced mass, and $V(r)>0$ is the e-h interaction with $r$ being the e-h distance.
We could also assume that $m_e\ll m_h$ and, hence, $\mu\sim m_e$ because electrons are usually much lighter than holes,
as depicted in Fig. \ref{fig1}, but this assumption is not necessary.
The exciton bound state energy is negative so that $E=-|E|$.
The wave vector consists of the radial ($k_r$) and tangential ($k_\varphi$) components, $k^2=k_r^2 + k_\varphi^2$.
The latter is quantized in terms of the magnetic quantum number $m=0,\pm 1, \pm 2...$ as $k_\varphi=m/r$.
The radial component is quantized using the Bohr-Sommerfeld quantization formula that in our case reads
\begin{equation}
 \label{quantum}
 \int\limits_{r_m}^{r_n} dr k_r = \pi(n+\gamma_M),
\end{equation}
where $r_m$, $r_n$ are the quasiclassical turning points, $n=0,1,2...$, is the radial quantum number,
and $0 \leq \gamma_M <1$ is the Maslov index.\cite{EPL2014goerbig}
We assume that the band structure is topologically trivial and set $\gamma_M=1/2$.
%Certainly, Eq. (\ref{quantum}) is not able to provide the exact values for energy levels, however, it is proven to be reasonable for modeling a relative shift of spectral lines upon an external parameter change.\cite{PRL2018trushin}
Furthermore, we simplify the problem considering solely $s$ excitons with zero angular momentum ($m=0$)
relevant for one-photon processes.
This simplification makes the region near $r=0$ quasiclassically accessible so that $r_m=0$ for any $n$.
Since $p\gg 2$ the first term dominates in Eq.~(\ref{total}) at $k<k_0$, whereas the second term skyrockets at $k>k_0$.
Hence, the wave vector as a function of $r$ can be approximately written as
\begin{eqnarray}
\nonumber &&  k_r\simeq k_0\,\theta\left[V(r)-E_{k_0}-|E|\right] +\sqrt{\frac{2\mu}{\hbar^2}\left(V(r)-|E|\right)}\\
 \label{momentum} 
 && \times\left\{\,\theta\left[V(r) -|E| \right] -\theta\left[V(r)-E_{k_0}-|E|\right] \,\right\},
\end{eqnarray}
where $E_{k_0}=\hbar^2 k_0^2 /(2\mu)$, and $\theta[x]$ is the Heaviside step function: $\theta[x] = 0$ at $x<0$, $\theta[x] = 1$ at $x>0$.
This approximation shrinks the set of band parameters to the reduced mass $\mu$ and
the size of the valence band flatness region $k_0$.
The major merit of the quasiclassical model is that it remains analytically tractable for 
any explicitly given e-h potential $V(r)$; however, we first focus on the Coulomb interaction
to qualitatively understand the influence of the band {\em flatness} on the exciton spectrum.

\subsection{Coulomb interaction}

\begin{figure}
\includegraphics[width=\columnwidth]{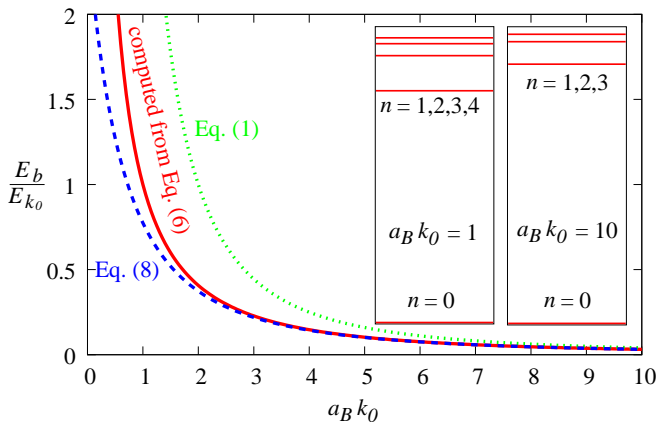}
\caption{\label{fig2} The exciton binding energy $E_b$ is plotted 
in units of $E_{k_0}=\frac{\hbar^2 k_0^2}{2\mu}$ as a function of $a_Bk_0$.
The latter is tuneable through the dielectric permittivity $\epsilon_\mathrm{env}$.
The red (solid) curve is given by the exact solution of Eq. (\ref{main-general}), the blue (dashed) curve is 
the approximate low-energy (or strong-screening) solution given by Eq. (\ref{binding1}), and 
the green (dotted) curve is the
hydrogenic solution given by Eq. (\ref{binding00}).
The inset shows the relative level spacing computed directly from Eq. (\ref{main-general}). The valence band {\em flatness} reduces the absolute value of the exciton binding energy and changes the relative level spacing in the spectrum.}
\end{figure}

Here, we set
\begin{equation}
\label{Coulomb}
V(r)=\frac{e^2}{\epsilon_\mathrm{env} r} 
\end{equation}
so that $r_n=e^2/(\epsilon_\mathrm{eff}|E|)$ and $r_m=0$ for $s$ states.
The integral in Eq.~(\ref{quantum}) can be taken easily, and the quantization condition reads
\begin{equation}
 \label{main-general}
\frac{\pi}{2} - \mathrm{arctan}\left(\frac{\sqrt{\varepsilon}}{2} - \frac{1}{2\sqrt{\varepsilon}}\right)
=\pi a_Bk_0\sqrt{\varepsilon}\left(n+\frac{1}{2}\right),
\end{equation}
where $\varepsilon=|E|/E_{k_0}$ is the dimensionless energy we are after.
This equation is used to calculate the exciton binding energy and excited spectral lines
depicted in Fig. \ref{fig2} by red color.

In the low-energy limit $\varepsilon\ll 1$ (equivalent to the rather strong screening $a_B k_0 \gg 1$),
the $\mathrm{arctan}$ term allows for a Taylor expansion as
\begin{equation}
\label{exp1}
 \mathrm{arctan}\left(\frac{\sqrt{\varepsilon}}{2} - \frac{1}{2\sqrt{\varepsilon}}\right) 
 =-\frac{\pi}{2} + 2\sqrt{\varepsilon} -o^{3/2}(\varepsilon).
\end{equation}
Combining Eqs. (\ref{main-general}) and (\ref{exp1}) we reproduce the Rydberg series shifted by $4/(\pi a_B k_0)$ as
$$
E_n=-\frac{2\hbar^2}{\mu a_B^2}\frac{1}{\left[2n + 1 + 4/(\pi a_B k_0)\right]^2}, \quad a_B k_0 \gg 1.
$$
The binding energy can be then written as
\begin{equation}
 \label{binding1}
 E_b=\frac{\pi^2 \hbar^2 k_0^2}{8 \mu}\frac{1}{\left(1 + \pi a_B k_0/4\right)^2},\quad a_B k_0 \gg 1.
\end{equation}
In contrast to Eq. (\ref{binding00}), the finite $k_0$ formally regularizes the divergence at $a_B\to 0$.
In the limit $a_B \to k_0^{-1}$,  Eq. (\ref{main-general}) allows for an exact solution given simply by $E_b=E_{k_0}$. 
The limit $a_B k_0  \ll  1$ does not make much sense as it implies that
the exciton size becomes comparable or even smaller than the lattice constant,
making the Wannier-Mott picture invalid.
It is, however, instructive to consider this limit from the formal point of view (see Appendix \ref{appA}). 

The valence band {\em flatness} influences the excited states as well.
Similar to the nonlocal screening and Berry curvature\cite{PRL2018trushin}
in 2D WSe$_2$ and WS$_2$,
the band {\em flatness} shifts the excited states towards the ground-state energy level,
reducing the level spacing between the binding energy and excited spectral lines.
Figure \ref{fig2} (inset) demonstrates how the energy spectrum changes once $a_B k_0$
is shifted from $10$ to $1$.

\subsection{Non-Coulomb interaction}

It is known for decades that the e-h interaction in thin semiconducting layers does not obey
Coulomb's law. \cite{rytova1965coulomb,keldysh1979coulombENG}
Some recent derivations\cite{PRB2014berghauser,PRB2014rodin,kylanpaa2015binding,PRB2015latini}
of the 2D e-h potential mostly rely on a finite extension of
the semiconductor layer (or layers) in the vertical direction and let this thickness vanish in a 2D limit.
The problem is that the resulting screening radius turns out to be proportional
to either the layer thickness\cite{PRB2014berghauser} or the
layer separation\cite{kylanpaa2015binding} and therefore formally vanishes in the 2D limit.
Since our semiconductor is assumed to be perfectly two-dimensional, this approach does not make much sense here. 
We therefore find it more practical to consider the 2D limit from the very beginning.
Here, we take into account the intrinsic polarizability\cite{cudazzo2011dielectric,PRB2014rodin} of a 2D semiconductor
that induces a charge density given by
$n_\mathrm{int} = \chi \delta(z)\nabla^2 \phi(\mathbf{r},z=0)$,
where $\phi(\mathbf{r},z)$ is the electrostatic potential,
$\chi$ is the in-plane polarizability, and $\delta(z)$ stands for the delta-function
providing an in-plane confinement.
In addition to that, we take into account
the out-of-plane polarizability (denoted as $\zeta$) due to the dielectric substrate.\cite{PRB2014rodin}
The charge density induced by the out-of-plane polarization then reads
\begin{equation}
n_\mathrm{env} = - \zeta \delta(z) \frac{\partial}{\partial z} \phi(\mathbf{r},z).
\end{equation}
The relation between $\zeta$ and $\epsilon_\mathrm{env}$  is given by
\begin{equation}
\epsilon_\mathrm{env}= 1+2\pi \zeta \equiv \frac{\epsilon_\mathrm{above} + \epsilon_\mathrm{below}}{2},
\end{equation}
where $\epsilon_\mathrm{above}=1$ and $\epsilon_\mathrm{below}=1+4\pi \zeta$
are the relative dielectric permittivities above and
below the 2D semiconductor layer, respectively. These settings imply
that our 2D semiconductor is placed on a dielectric substrate without cover.
The two parameters $\epsilon_\mathrm{env}$ and $\zeta$ do describe the same 
physical effect and their roles are interchangeable.

The external charge is pointlike, and the corresponding 
charge density is given by $n_\mathrm{ext} = e\delta(z)\delta(\mathbf{r})$.
The total charge density is
$n_\mathrm{tot}=n_\mathrm{ext}+n_\mathrm{int}+n_\mathrm{env}$, and the potential $\phi(\mathbf{r},z)$ 
must satisfy the Poisson equation 
\begin{equation}
\label{poisson}
 \nabla^2 \phi(\mathbf{r},z) = -4\pi n_\mathrm{tot}.
\end{equation}
This equation has been solved by Cudazzo {\it et al.}, \cite{cudazzo2011dielectric}
in the limit $\zeta=0$ (no substrate).
Here, we solve it at $\zeta\neq 0$.
We employ the Fourier transformation
$\phi(\mathbf{r},z)=\int d^2 q \tilde\phi(\mathbf{q},z) 
\mathrm{e}^{i \mathbf{q}\cdot\mathbf{r}}$
and rewrite Eq. (\ref{poisson}) as
\begin{eqnarray}
\label{poisson2}
&& \left(\frac{\partial^2}{\partial z^2} - q^2\right)\tilde\phi(\mathbf{q},z)  = \\
\nonumber && -4\pi  \left[e \delta(z) - \chi \delta(z) q^2 \tilde\phi(\mathbf{q},z=0)
-  \zeta \delta(z) \frac{\partial}{\partial z} \tilde\phi(\mathbf{q},z) \right].
\end{eqnarray}
The solution of Eq. (\ref{poisson2}) is given by
\begin{equation}
\label{ansatz}
\tilde\phi(\mathbf{q},z) = u_q \mathrm{e}^{-q |z|},
\end{equation}
where $u_q=2\pi e/(q \epsilon_q)$ with 
$\epsilon_q=\epsilon_\mathrm{env}(1 + r_s q)$, $r_s=2\pi\chi/\epsilon_\mathrm{env}$.

The resulting in-plane real-space interaction energy reads
\begin{equation}
\label{Cudazzo}
V(r)=\frac{\pi e^2}{2\epsilon_\mathrm{env} r_s} 
\left[H_0\left(\frac{r}{r_s}\right) - Y_0\left(\frac{r}{r_s}\right)\right],
\end{equation}
where $H_0$ is the Struve function, and $Y_0$ is the Bessel function of the second kind.
At $r\gg r_s$, Eqs. (\ref{Cudazzo}) and (\ref{Coulomb}) just coincide,
but in the opposite limit $r \ll r_s$ the interaction reads
\begin{equation}
\label{Cudazzo-apprx}
V(r)\simeq \frac{e^2}{\epsilon_\mathrm{env} r_s} 
\left[\ln\left(\frac{2r_s}{r}\right) - C \right],\quad r\ll r_s,
\end{equation}
where $C= 0.5772$ is the Euler's constant.

\begin{figure}
\includegraphics[width=\columnwidth]{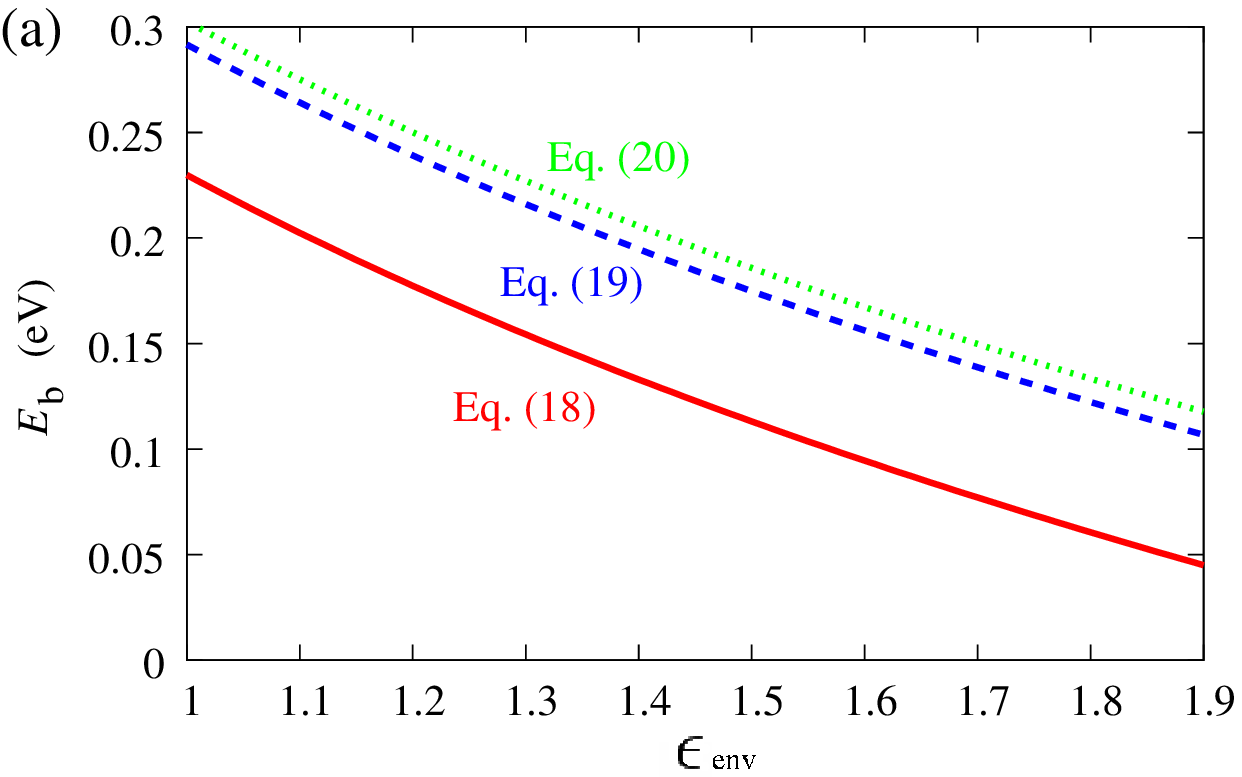}
\includegraphics[width=\columnwidth]{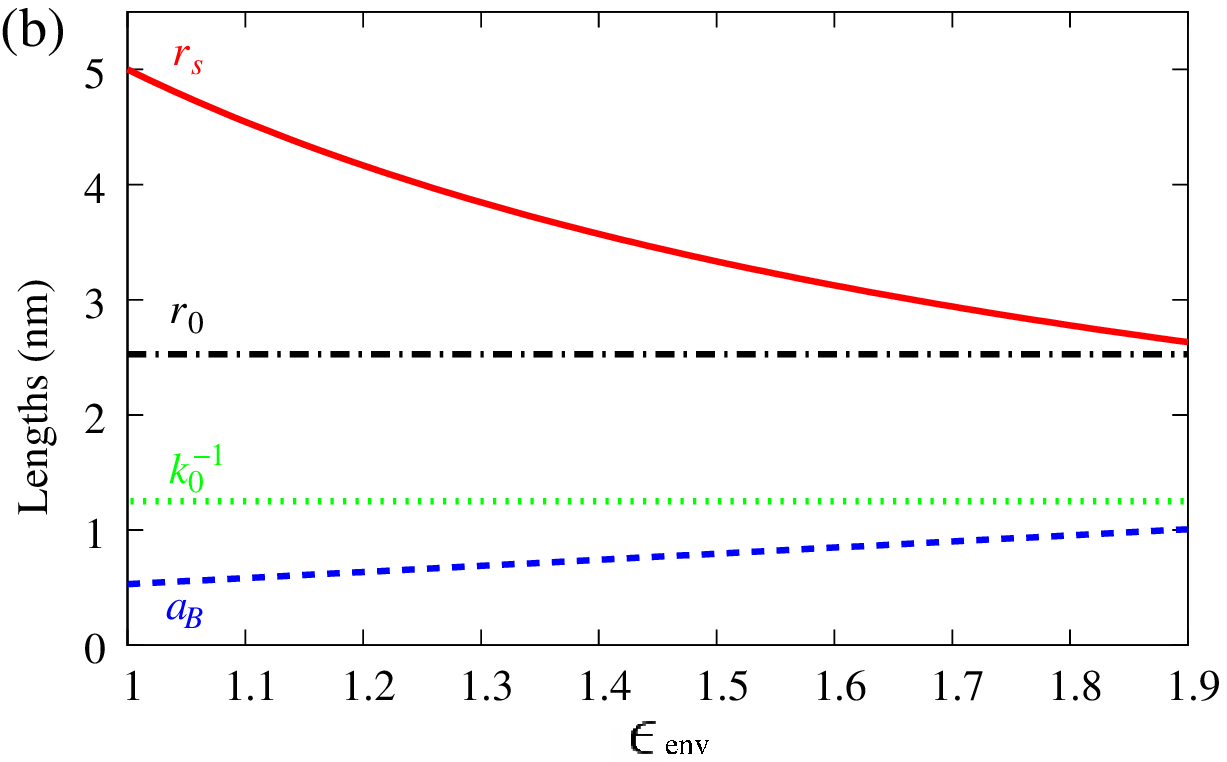}
\caption{\label{fig3} (a) The exciton  binding energy $E_b$ as a function of the dielectric permittivity of
the media surrounding the host 2D semiconductor. We take $\mu=0.1 m_0$ (here, $m_0$ is the free electron mass),
$k_0=0.08\,\mathring{\mathrm{A}}^{-1}$, and $r_s=5$ nm in vacuum with $\epsilon_\mathrm{env}=1$
(see the main text for discussion).
The red (solid) line represents the main result given by Eq. (\ref{binding11}),
the blue (dashed) line shows what happens at 
$k_0 \sqrt{a_Br_s/2}\gg 1$ in accordance with Eq. (\ref{binding10}),
and the green (dotted) line is given by Eq. (\ref{binding-ultimate}), corresponding to 
$k_0 \sqrt{a_Br_s/2}\ll 1$.
For realistic parameters used here $k_0 \sqrt{a_Br_s/2} \sim 1$, so that
both approximations (\ref{binding10}) and (\ref{binding-ultimate}) substantially overestimate 
the binding energy.
(b) The characteristic lengths as functions of the relative permittivity $\epsilon_\mathrm{env}$:
$r_0$ is the ground-state exciton radius defined in the main text, 
$a_B=\epsilon_\mathrm{env}\hbar^2/(\mu e^2)$ is the effective Bohr radius,
$k_0^{-1}$ is the band {\em flatness} parameter inverted, and 
$r_s=2\pi \chi/\epsilon_\mathrm{env}$ is the screening radius. The model is applicable as long as $r_0$
is substantially smaller than $r_s$,
so that the interaction is given by Eq. (\ref{Cudazzo-apprx}) rather than by the Coulomb potential.}
\end{figure}

The quasiclassical ground-state radius of a tightly bound exciton is smaller than $r_s$,
%the electron-hole interaction is given by Eq. (\ref{Cudazzo-apprx}) rather than by Eq. (\ref{Coulomb}).
%that allows for an explicit experession for the binding energy of such excitons.
which justifies the applicability of Eq. (\ref{Cudazzo-apprx}) for the binding energy calculations.
We rewrite the quantization condition (\ref{quantum}) for the ground state $n=0$ as
\begin{equation}
\label{quantum2}
 k_0 r'_0 + \sqrt{\frac{2}{a_B r_s}}\int\limits_{r'_0}^{r_0}dr \sqrt{\ln\left(\frac{r_0}{r}\right)}  = \frac{\pi}{2},
\end{equation}
where $r_0=2r_s e^{-\frac{\epsilon_\mathrm{env}r_s}{e^2}E_b - C}$ and
$r'_0=r_0 e^{-\frac{\epsilon_\mathrm{env}r_s}{e^2}E_{k_0}}$.
The integral can be taken as
\begin{equation}
\label{integral}
\int\limits_{r'_0}^{r_0}dr \sqrt{\ln\left(\frac{r_0}{r}\right)}=
\frac{\sqrt{\pi}}{2}r_0\,\mathrm{erf}\left(\sqrt{\ln\frac{r_0}{r'_0}}\right)-r'_0\sqrt{\ln\frac{r_0}{r'_0}},
\end{equation}
where $\mathrm{erf}$ is the error function.
Solving Eq. (\ref{quantum2}) with respect to $E_b$  we obtain the main result of this work:
\begin{eqnarray}
\nonumber &&
 E_b= \frac{e^2}{\epsilon_\mathrm{env} r_s} \left(
 \ln\left\{r_s k_0  e^{-\frac{a_Br_sk_0^2}{2}} \right.\right.\\
\nonumber && \left. \left.
 + \sqrt{\frac{2r_s}{a_B}}\left[\frac{\sqrt{\pi}}{2}\mathrm{erf}\left(k_0\sqrt{\frac{a_B r_s}{2}}\right)
 - k_0\sqrt{\frac{a_B r_s}{2}} e^{-\frac{a_Br_sk_0^2}{2}}\right] \right\} \right.\\
&& \left. - C - \ln \frac{\pi}{4} \right).
 \label{binding11}
\end{eqnarray}
In contrast to the hydrogenic model, Eq. (\ref{binding11}) suggests a weak logarithmic dependence of
the binding energy on $\epsilon_\mathrm{env}$. This dependence is shown in Fig. \ref{fig3}(a) by the red (solid) line.
Moreover, the exciton radius $r_0$ does
not depend on $\epsilon_\mathrm{env}$ at all. Hence, a tightly bound 2D exciton is not much influenced
by the environment once its radius $r_0$ is substantially smaller than the intrinsic screening radius $r_s$.
In contrast to bulk semiconductors, the exciton binding energy value does not tell us much about  
its radius, as has already been noted by Cudazzo {\em et al.}\cite{PRL2016cudazzo}

The binding energy value is governed by the dimensionless parameter $k_0\sqrt{a_Br_s/2}$,
which combines three length scales characterizing 
the band {\em flatness} ($k_0^{-1}$), the Coulomb interaction strength ($a_B$),
and its screening due to intrinsic polarizability ($r_s$).
The typical values for these lengths are shown together with the quasiclassical exciton radius $r_0$
in Fig. \ref{fig3}(b) as functions of the relative dielectric permittivity $\epsilon_\mathrm{env}$.
Note that $k_0\sqrt{a_Br_s/2}$ does not depend on $\epsilon_\mathrm{env}$ because
$a_B\propto \epsilon_\mathrm{env}$, $r_s\propto 1/\epsilon_\mathrm{env}$, and $k_0=\mathrm{const}$.
At $k_0 \sqrt{a_Br_s/2}\gg 1$, we can estimate the binding energy as
\begin{equation}
 E_b= \frac{e^2}{\epsilon_\mathrm{env} r_s} \left(\ln\sqrt{\frac{8r_s}{\pi a_B}} -C\right),\quad k_0 \sqrt{a_Br_s/2}\gg 1.
 \label{binding10}
\end{equation}
This is a known relation formally obtained for tightly bound 2D excitons a long time ago.\cite{keldysh1997excitons}
The binding energy (\ref{binding10}) still diverges at $a_B\to 0$, 
but, in contrast to the hydrogenic binding energy (\ref{binding00}), the divergence
that results in the lower values shown in Fig. \ref{fig3}(a) is logarithmic.
Once $k_0$ is finite, the binding energy given by Eq. (\ref{binding11}) does not diverge at $a_B\to 0$.
Hence, the valence band {\em flatness} indeed limits the binding energy from above, and 
its limiting value can be written as
\begin{equation}
 E_b= \frac{e^2}{\epsilon_\mathrm{env} r_s} \left(\ln \frac{4k_0 r_s}{\pi}  -C \right),\quad k_0 \sqrt{a_Br_s/2}\ll 1.
 \label{binding-ultimate}
\end{equation} 
Equations (\ref{binding10}) and (\ref{binding-ultimate}) could also be derived starting from
finite $p$ (see Appendix \ref{appA} for details).

Figure \ref{fig3}(a) shows the binding energy behavior described 
by Eqs. (\ref{binding11}), (\ref{binding10}), and (\ref{binding-ultimate}).
In the next subsection, we find that the most relevant regime corresponds to $k_0 \sqrt{a_Br_s/2}\sim 1$,
so that one should utilize the most complicated equation of those three to obtain a realistic estimation of $E_b$.

\subsection{Particular examples}

The most studied 2D semiconductors are probably 2D transition metal dichalcogenides (2DTMDCs)
characterized by the {\em ab initio} calculated\cite{berkelbach2013theory}
$r_s \sim 4$ -- $5\,\mathrm{nm}$ and $\mu\sim 0.1$ -- $0.2 m_0$ (in vacuum). 
$E_b$ and $a_B$ depend on the dielectric permittivity of the environment 
(see Refs. \onlinecite{kylanpaa2015binding} and \onlinecite{stier2016probing} for calculations and measurements, respectively).
In the most common 2DTMDCs (like MoS$_2$, WS$_2$, and WSe$_2$) the valence band {\em flatness} is negligible.
However, by playing with different transition metals and lattice deformations one can bring
the band structure to a less boring shape.
In particular, monolayer 1T--TiSe$_2$ in a distorted phase demonstrates a nearly flat valence band\cite{wei2017TiSe2,singh2017TiSe2,singh2018TiSe2} with well-defined $k_0=0.0785\,\mathring{\mathrm{A}}^{-1}$
(see the Supplemental Information in Ref. \onlinecite{singh2018TiSe2}).
The conduction band remains rather parabolic at the scale of $k_0$ resembling
the band structure shown in Fig. \ref{fig1}.
An additional effect of the valence band inversion can be absorbed into the reduced effective mass in the
first term of Eq. (\ref{total}). 
Combining these parameters together we find that $k_0 \sqrt{a_Br_s/2}\sim 1$; hence
the band {\em flatness} is important, and Eq. (\ref{binding11}) applies.
Another promising 2D semiconductor with a flat valence band is the distorted 2D 1T--WTe$_2$
recently fabricated by using molecular beam epitaxy on a bilayer graphene substrate.\cite{tang2017quantum}
Among the 2DTMDC family members, WTe$_2$ is the only one 
for which the distorted 1T--phase (also known as the 1T$'$--phase) 
is most energetically favored.\cite{qian2014quantum}
The valence band turns out to be perfectly flat with $k_0$ of the order of $0.1\,\mathring{\mathrm{A}}^{-1}$.
However, the conduction band minimum is shifted away from the center of the first Brillouin zone,
resulting in a slightly indirect band gap.
Anyway, both materials (monolayer 1T$'$--TiSe$_2$ and 1T$'$--WTe$_2$)
possess the band gap size of less than $100$ meV
comparable with the exciton binding energy estimated from our model.
Figure \ref{fig3} suggests that the binding energy may reach the size
of the band gap at $\epsilon_\mathrm{env} \lesssim 2$,
corresponding to a substrate with $\epsilon_\mathrm{below} \lesssim 3$.
There is already evidence for an excitonic insulator phase in bulk 1T--TiSe$_2$ (see Ref. \onlinecite{PRL2007cercellier}),
and any observation of such a phase transition in the 2D limit could be very helpful
for understanding the underlying mechanism.

The present model is also related to the tuneable excitons in biased bilayer graphene.
It has been theoretically shown a long time ago\cite{park2010tunable}
that the optical response of this system is dominated by bound e-h states.
The electron and hole dispersions in biased bilayer graphene are far from being parabolic
and the excitons are tightly bound:
The characteristic $k_0$ is about $0.02\,\mathring{\mathrm{A}}^{-1}$,
and the excitonic ground-state radius $r_0$ is about $10$ nm, which results 
in $k_0 r_0\sim 1$. The exciton spectrum must be therefore strongly influenced 
by the non-parabolicity of the carrier dispersion. 
The {\em ab initio} study\cite{park2010tunable} accounts for this non-parabolicity automatically
and indeed predicts strong deviations from a 2D hydrogenic model
(see also recent measurements\cite{ju2017tunable}).
Note that the present model is not directly applicable to biased bilayer graphene
because of the deep local minima in the carrier dispersions.\cite{PRA2015shvartsman}

Another candidate could be thin samples of Ga$_2$Se$_2$ (or Ga$_2$S$_2$) with a perfectly flat valence band 
characterized by $k_0\sim 0.1\,\mathring{\mathrm{A}}^{-1}$ for the thickness comprising eight (or six for Ga$_2$S$_2$)
tetralayers.\cite{rybkovskiy2014Ga2Se2,PCCP2013ma} The {\em flatness} and subsequent valence band inversion 
occurring with a further decrease of sample thickness seems to be responsible 
for the strong reduction of the excitonic absorption observed recently\cite{budweg2018suppression}
in few-layer Ga$_2$Se$_2$.
Other members of this family include Ga$_2$Te$_2$ and In$_2$Se$_2$.\cite{zolyomi2013Ga2Se2,rybkovskiy2014Ga2Se2}

\section{Conclusion}

To conclude, this study shows what to expect once 2D tightly bound excitons become so small
that the effective-mass approximation is not applicable anymore,
but the exciton size remains much larger than the lattice constant,
maintaining the Wannier-Mott character.
To quantify the effect, we have introduced the characteristic wave vector $k_0$ 
separating two regions in the first Brillouin zone:
At $k\ll k_0$ the electron and hole bands can be well described
in terms of the corresponding effective masses,
whereas at $k \gtrsim k_0$ the quadratic approximations do not match the actual dispersions.
The exciton behavior is then determined by the ratio between $k_0^{-1}$ and
other characteristic lengths, such as intrinsic 2D polarizability and effective Bohr radius.
Assuming that the hole energy rapidly increases at $k>k_0$
we find that the exciton binding energy is reduced as compared to the limiting case of $k_0^{-1}=0$.
Formally speaking, the finite value of $k_0^{-1}$ regularizes the formal divergence at the vanishing effective Bohr radius.
In other words, the sudden change of dispersion at $k>k_0$ prevents the exciton energy from skyrocketing into the eV range
even though the hosting 2D semiconductor is suspended in vacuum providing no external screening.
The model proposed can be used for exciton binding energy estimations in several narrow-gap 2D semiconductors
with nearly flat valence bands where a phase transition to the excitonic insulator is expected.

\acknowledgments

This work has been supported by the Director's Senior Research Fellowship 
from the Centre for Advanced 2D Materials at the National University of Singapore
(NRF Medium Sized Centre Programme R-723-000-001-281).
I thank Vitor Pereira, Giovanni Vignale, Alex Rodin, and Keian Noori for discussions.

\appendix

\section{Heavy carriers}
\label{appA}
 
Here, we assume that $p$ (the {\em flatness} degree) is not necessarily much larger than $1$,
but the carriers are so heavy that the first term in Eq.~(\ref{total}) can be neglected.
If that term is retained, then such a regime corresponds to the extreme limit $a_B k_0 \ll 1$ 
and is never realized in conventional 2D semiconductors. 
It is, however, instructive to consider this regime for the sake of completeness.
After all, the conduction band could also be flat, making this model feasible.

The quantization condition in the case of Coulomb interactions reads
\begin{equation}
 \label{arb-p}
 k_0\int\limits_0^{r_n} dr \left(-\frac{|E|}{\beta} + \frac{e^2}{\epsilon_\mathrm{env} \beta r}\right)^{\frac{1}{p}}= \pi\left(n+\frac{1}{2}\right),
\end{equation}
where $r_n=e^2/(\epsilon_\mathrm{env} |E|)$. Integrating Eq.~(\ref{arb-p}) we obtain the following spectrum
\begin{equation}
\label{arb-p-result}
 E_n=  -\frac{1}{\left[\left(n+1/2\right) p\sin\left(\frac{\pi}{p}\right)\right]^{\frac{p}{p-1}} \beta^{\frac{1}{p-1}}} \left(\frac{e^2 k_0}{\epsilon_\mathrm{env}}\right)^{\frac{p}{p-1}}.
\end{equation}
In the conventional case ($p=2$) the spectrum obeys the standard Rydberg $1/(2n+1)^2$-series. 
The spectrum approaches the unconventional $1/(2n+1)$-series
with the binding energy $E_b=2e^2k_0/(\pi \epsilon_\mathrm{env})$
as the {\em flatness} increases ($p\to \infty$).
The exciton ground-state radius is then determined solely by $k_0$ as $r_0=\pi/(2k_0)$.

In the case of $V(r)$ given by Eq. (\ref{Cudazzo-apprx}), the quantization condition reads
\begin{equation}
 \label{arb-p-nc}
 k_0 \left(\frac{e^2}{\epsilon_\mathrm{env} \beta r_s} \right)^{\frac{1}{p}}
 \int\limits_0^{r_n} dr \left(\ln\frac{r_n}{r}\right)^{\frac{1}{p}}= \pi\left(n+\frac{1}{2}\right),
\end{equation}
where $r_n=2r_s e^{-\frac{\epsilon_\mathrm{env}r_s}{e^2}|E| - C}$ with $r_n \ll r_s$.
The corresponding spectrum is given by
\begin{eqnarray}
\label{arb-p-result-nc}
&& E_n=  -\frac{e^2}{\epsilon_\mathrm{env} r_s} \\
\nonumber && \times \left\{ \ln\left[\frac{4k_0 r_s \Gamma(1+\frac{1}{p})}{\pi(2n+1)}
 \left(\frac{e^2}{\epsilon_\mathrm{env} \beta r_s} \right)^{\frac{1}{p}}\right] -C \right\},
\end{eqnarray}
where $\Gamma$ is the Euler's gamma function.

Using Eq. (\ref{arb-p-result-nc}) it is possible to double-check our previous expressions for the binding energy.
In the limit $p\to\infty$ we have 
\begin{equation}
 E_n=  -\frac{e^2}{\epsilon_\mathrm{env} r_s}
\left\{ \ln\left[\frac{4k_0 r_s}{\pi(2n+1)}\right] -C \right\},
\end{equation}
so that the binding energy is given by Eq. (\ref{binding-ultimate}).
In the limit $p=2$ the spectrum reads
\begin{equation}
 E_n=  -\frac{e^2}{\epsilon_\mathrm{env} r_s}
\left\{\ln\left[\frac{2k_0 r_s}{\sqrt{\pi}(2n+1)}
 \sqrt{\frac{e^2}{\epsilon_\mathrm{env} \beta r_s}}\right] -C \right\}.
\end{equation}
Assuming $\beta=\hbar^2k_0^2/(2\mu)$, we arrive at Eq. (\ref{binding10}) for the binding energy.

\end{document}